\documentclass[conference]{IEEEtran}
\IEEEoverridecommandlockouts
\usepackage{cite}
\usepackage{amsmath,amssymb,amsfonts}
\usepackage{algorithmic}
\usepackage{graphicx}
\usepackage{textcomp}
\usepackage{xcolor}
\usepackage{float}
\usepackage{caption}
\usepackage{enumitem}

\usepackage[linesnumbered,lined,boxed,commentsnumbered,ruled,vlined]{algorithm2e}

\def\BibTeX{{\rm B\kern-.05em{\sc i\kern-.025em b}\kern-.08em
    T\kern-.1667em\lower.7ex\hbox{E}\kern-.125emX}}

\def\BibTeX{{\rm B\kern-.05em{\sc i\kern-.025em b}\kern-.08em
    T\kern-.1667em\lower.7ex\hbox{E}\kern-.125emX}}
\begin{document}
\IEEEoverridecommandlockouts
\IEEEpubid{\makebox[\columnwidth]{978-1-7281-3455-0/19/\$31.00 \copyright 2019 IEEE \hfill} \hspace{\columnsep}\makebox[\columnwidth]{ }}
\hbadness=99999

\title{From Blockchain to Hashgraph: Distributed Ledger Technologies in the Wild
}

\author{\IEEEauthorblockN{Zuhaib Akhtar}
\IEEEauthorblockA{
\textit{Department of Computer Engineering}\\
Aligarh Muslim University \\
Aligarh, India\\
akhtarzuhaib@gmail.com
}
}

\hyphenchar\font=-1
\maketitle

\begin{abstract}
With the introduction of the term blockchain in 2008, it's interest has been increasing in the community since the idea was coined. The reason for this interest is because it provides anonymity, security and integrity without any central third party organisation in control of data and transaction. It has attracted huge interest in research areas due to its advances in various platforms, limitations and challenges. There are various Distributed Ledger Technologies that demonstrates their special features which overcome limitations of other platforms. However, implementations of various distributed ledger technologies differ substantially based on their data structures, consensus protocol and fault tolerant among others. Due to these variations, they have a quite different cost, performance, latency and security. In this paper, working and in-depth comparison of major distributed ledger technologies including their special features, strengths and weaknesses is presented and discussed by identifying various criteria.  
\end{abstract}

\begin{IEEEkeywords}
Distributed ledger technology, Blockchain, Sidechain, Tangle, Hashgraph
\end{IEEEkeywords}

\section{INTRODUCTION}
The advent of protocols like TCP/IP and evolution of the internet led to the development of centralised systems. These systems are used in various domains. One such domain is centralised transactions. Currency transactions between companies or persons are generally centralised and these centralised servers are controlled by third party organisations. As they are providing services, a transaction fee is charged by them. All the information and data are managed and controlled by them, rather than the two entities that were involved in the transaction. With the development of blockchain technology, this issue has been resolved. Its goal is to create decentralised network, where there is no control of the third party over data and transactions. It is a distributed ledger technology (DLT) which stores data and transaction in the network itself rather than in any centralised system. It guarantees safety, privacy, integrity and transparency to the user [1]. People can transfer the cryptocurrency to anyone and even to those who do not have a bank account.
However, blockchain has faced criticism for its scalability and cost. Therefore, new Directed Acyclic Graph (DAG) based DLTs have been introduced such as hashgraph. It promises to remove most of the shortcomings of blockchain based DLTs.\\ 
In this study, working of major DLTs and systematic comparison between various DLTs are presented. The objective behind this effort is to define and compare the main properties and crucial advantage of different technologies, and identify the current state of DLTs. Based on this research, a set of quality criteria has been identified which helps to compare their special features and shows the sequential progress of DLTs over the years. 

\section{Major distributed ledger technologies review}
Major goal of DLT is to bring trust into the system in which there is no central authority which can be trusted. In such a system, there is no mutual trust between users. At its core, DLT are data structures including functions to handle them. Each DLT has its own data model and mechanism, but all are based on three technologies i.e., distributed peer-to-peer network, public key cryptography and consensus algorithm. As in distributed environment, no one can be trusted, public key cryptography brings security and digital identity for every system in the network. Each participant has a public and a private key pair which helps in recording transactions in Distributed Ledger (DL). Members enforce control on objects managed by DL with the help of digital identity. A distributed peer-to-peer is meant to avoid single point of failure, scale easily and effectively in the  network,  and to avoid hostile takeover by a member or a group in the network. Consensus protocol is the heart of any DL. It brings trust into the system where one member cannot trust the other i.e., all members agree on the one and only one version of truth. A consensus algorithm must be deterministic, fair, secure and fast. It must tolerate byzantine behavior [2]. In byzantine consensus, correct processes or systems agree on a value and eventually entire system agrees on it even if some systems are malicious. Following sub-sections discusses major DLTs along with their features, strengths and weaknesses. 

\subsection{Blockchain}
Blockchain is immutable, distributed and decentralised ledger which stores transactions in the form of blocks. The blocks are appended as time passes and is accessible to all members of the network. Blocks are connected to each other in the form of linked list. The major advantage is that it makes difficult for any member to tamper with the contents of the block as any change in it will break the chain of blocks. This is because every block contains hash of previous block. The elements of blockchain are described as follows:
\begin{itemize}
 \item Transaction is an information which is digitally signed by a member in the network. Group of transactions are known as blocks and they are appended to blockchain upon validation.
  \item A blockchain is a ledger which stores all the blocks that are created in the network. It utilises hash of previous block which connects all the blocks together, which is comparable to linked list.
  \item A consensus algorithm decides which block will be added to the blockchain.
\end{itemize}
Bitcoin and ethereum are two most common applications of blockchain. The consensus mechanism used in bitcoin is proof-of-work (POW). The core idea of this algorithm is to give rights (to add a block in the blockchain) to a member in the network that solves tough mathematical problem first. This tough mathematical problem is inverse hash calculation which is done by specialised hardware. Whosoever solves first gets the right to add next block to the blockchain. This is known as mining. However, these specialised hardware consumes a lot of power and are costly. Ethereum solves this issue by using proof-of-stake consensus algorithm. It replaced complex inverse calculation with an alternative approach which involves stake (cryptocurrency) of user. Longer a user holds stake, more rights it will get. Hence, it saves problem of wastage of resources. Some features of blockchain are given as follows:

\begin{itemize}
  \item \textbf{Implementations:} There is large number of implementation of blockchain based technologies and it is increasing. These technologies target different domain. However, main goal of blockchain is to create a DL that allows all members in the network to agree on and share the same version of truth.
  \item \textbf{Smart contracts:} Executable program that resides in blockchain and when some specific conditions are met, they get executed. They are not executed until their invoking transaction gets into the block.
  \item \textbf{Miners:} Mining provides incentives to that member who mines the new block i.e., one who gets the right to add a new block to blockchain.
\end{itemize}

\subsection{Sidechain}
Sidechain combines two different blockchain architectures to remove existing shortcomings of blockchain in terms of performance, privacy and security [3]. 
`Main' blockchain processes all the global requests of members and `sidechain' can be used to manage local requests. Even if sidechain gets compromised, blockchain can function on its own.\\
As number of nodes (members) grow in blockchain, it becomes difficult to achieve consensus in the network. Sidechain can be seen as dividing the main blockchain into segments. Sidechains are sub-networks and any request submitted to it is processed locally. Hence, rather than having single large main blockchain, it groups blockchain into number of sub-blockchains and a main blockchain. Each sidechain in the network can be used by a company or members with common interests. Digital assets can be moved back and forth between sidechains at a fixed exchange rate by the two-way peg mechanism. Sidechain can also be used to hide data from others. Sidechain is connected to the main blockchain through special nodes which are called validators whose work is to validate transactions that are occurring locally. It makes sidechain independent from the main blockchain.\\
Sidechain extends all the properties of blockchain but solves its major issues. Important features of sidechain are given as follows:
\begin{itemize}
\item \textbf{Scalability:} It resolves scalability issue of blockchain by creating sub-networks.
\item \textbf{Privacy:} It also solves privacy issues by imposing constraint over who can access which part of data in the network.
\end{itemize}

\subsection{Tangle}
Tangle if one of the platforms that is very well suited for Internet of Things (IOT). It was developed by IOTA. Tangle uses DAG based data structure. Each vertices in the DAG is called site and edges between sites corresponds to transaction approval.\\ 
IOTA was developed to solve many problems related to blockchain. It does not have miners, concept of blocks or transaction fees. Removing transaction fees is important in IOT, where M2M micro-transaction is expected. This is because transaction fee can be greater than the transaction itself. As there are no miners, any member that issues a new transaction, as a new site x, then it must select two existing sites y and z and validates those transactions by doing small POW. POW work algorithm in IOTA is called hashcard which works well in IOT devices that generally have low computation power. In IOTA, there is no upper bound on rate of transaction by the network.\\
Initial vertices are called genesis that holds all the crypto-currency of the system. Sites that are not yet validated are called tips. Node (a member) can choose any site for validation but honest nodes generally follows tip selection procedure. It starts from genesis and does weighted random walk till tip is reached. Weight of the site is dependent upon the number of validation it received previously.\\
Currently, IOTA validates transaction using coordinator node. More precisely, it issues a transaction every few minutes which is known as milestone. All the transactions referenced by milestone are validated. It is a temporary solution and sooner or later it will use a distributed solution known as Markov-chain Monte-Carlo algorithm.\\ 
Tangle proposes an Markov-chain Monte-Carlo algorithm which probabilistically checks validation of transaction. It runs tip selection procedure x times. It checks, for a given transaction, how many of the selected tip references it. Let's say these tips are y in number. Then, transaction is validated with the confidence y/x.
Some features of tangle are as follows:

\begin{itemize}
\item \textbf{Scalability:} As all members in the network verify transactions, increase in the number of  members results in faster validation. These members can also validate transactions in parallel.
\item \textbf{Micro-transactions or Transaction fees:} IOTA combined miner and validator into one role i.e., if one wants to issue a transaction, then it should support network by validating two transactions using small POW. Therefore, instead of paying miner additional fee for validating a transaction, member uses its own computing power to validate it. 
\item \textbf{Decentralisation:} As there are no special validators of transactions like miners in the network, therefore, there isn't any problem  of concentration of resources or computing power unlike bitcoin.
\item \textbf{Quantum resistant:} To generate public address, IOTA requires use of Winternitz One Time Signature scheme which is supposed to be quantum resistant. It is advised in IOTA's documentation that one should use generated address once as parts of private key might get disclosed.
\end{itemize}

\subsection{Hashgraph}

Hashgraph was developed by the co-founder and chief technology officer of Swirlds, Leemon Baird in 2016. It is a platform which can work in a malicious environment and provides distributed consensus. Swirlds is a permissioned network i.e., only authorised members can join the network and every member know about all the members in the network. \\
At an abstract level, hashgraph can be considered a structure which has columns and each member is represented by a column in the network. All the columns have many vertices. Each of the vertex is called an event. A user in the network basically performs two actions; (i) At any time user can create an event and submit it, (ii) User randomly picks a member in the network and gossips all the information that it know i.e., it sends the information to the member about the creation of event. The distribution of events takes place with the help of gossip-about-gossip protocol. These events store four different types of information in its data structure:
\begin{itemize}
\item Zero or more transactions submitted by the user
\item Timestamp or time at which event was created
\item Hash of the previous event created by the user 
\item Hash of the another user's event sending the gossip
\end{itemize}
User then places his digital signature and gossips about this event. The information, in addition to transactions in an event, serves various purposes. Incorporation of two different hashes allow members to know the history of transactions i.e., from where transaction was originated and where it was directed (to whom it was gossiped). Hashgraph uses this information to build DAG of events. DAG is updated as members gossips in the network. Signing an event helps in identifying the creator of event and no-one can tamper with that event.
It is pretty clear why protocol for distributing of event is known as gossip-about-gossip. Member sends not only what he know (gossip) but when other members learned it (gossip-about-gossip). This protocol has logarithmic running time. Suppose, at time t=0 a member creates an event, at time t=1 that member sends that event to randomly chosen other member. Hence, two members know that event at time t=1. Similarly,  those two members shares that event with two other members in the network at time  t=2. Hence, four members know about that event at time  t=2. As we can see, at time t=3, eight members will know about that event. Hence if there are `n' members in the network, it will take t=Log(n) for that event to spread in the entire network. Hence, gossip spreads in the network at an exponential rate.\\
Subsequent subsection discusses hashgraph consensus algorithm and its assumptions.
\subsubsection{Terminologies used in Hashgraph Consensus Algorithm}
\begin{itemize}
\vspace{-4.3mm}
\item Supermajority is a number which is greater than 2/3rd of any positive integer.
\item An event A is parent/ancestor of event B if there is an edge from A to B.
\item An event A is self-parent/self-ancestor of event B if A and B are created by same node.
\item An event A can see event B if B is an ancestor of A and creator of B does not have two ancestors of B such that these two events are not ancestors of each other.
\item An event A can strongly see B, if B is seen by more than supermajority of distinct nodes in the network.
\item First event of every member in a round (will be discussed in subsequent section) is termed as witness.
\end{itemize}

\subsubsection{Assumptions of Hashgraph Consensus Algorithm}

\begin{itemize}
\item number of members or nodes in the network is fixed and identity of all nodes in the network is known.
\item Number of malicious nodes in the network is less than 1/3rd of the total members in the network.
\item Signature of members cannot be forged and hash function used is resistant to collision.
\item Messages transfer between members of the network is asynchronous. This means that adversary can delay the message but it will eventually reach the whole network.
\end{itemize}

\subsubsection{Hashgraph Consensus Algorithm}

\begin{algorithm}
\textbf{Run two loops in parallel\;}
    \While{True}{
         Sync all known events to a random member\;
    }
    \While{True}{
        receive a sync\;
        create a new event\;
        \textbf{call} divideRounds\;
        \textbf{call} decideFame\;
        \textbf{call} findOrder\;
    }
\caption{Hashgraph Consensus Algorithm}
\end{algorithm}
Hashgraph consensus algorithm run two loops in parallel. First loop gossips with other members in the network, sending entire history of transactions. Second loop has three sub-protocols. Node in the network wait for the gossip from any other node. After it receives gossip (sync), it takes hash of that gossip, hash of its previous event, some new transaction and places timestamp of that event. Then event is signed by its creator. After that, node runs three sub-procedures to achieve consensus on the order of events. These are given below:

\begin{itemize}
\item \textbf{DecideRound:} First event of every node is assigned round number 1. After that if any event can strongly see supermajority of round r witnesses, then it is assigned round number r+1. Otherwise, it will be r.
\item \textbf{DecideFame:} A witness in round r can be decided famous. If witness in round r+1 can see witness in round r, it votes YES. Otherwise, it votes NO. Each of the witness in round r+1 votes for witness in round r in similar fashion. These votes are counted by witness in round r+2. If total count of YES votes exceeds supermajority, then it is decided that witness in round r is famous and election is over. If supermajority of votes are NO, then witness in round r is not famous. If witness in round r+2 cannot decide election, then other witness count votes to decide election. Witness in round r+2 count YES/NO vote of witness in round r+1, if it strongly sees witness in round r+1. It might be possible that none of the witness in round r+2 can decide election. In such case, witness in succeeding round will decide election. There might be a case where election might not end. To counter this problem, coin round is conducted. Witness merely vote when collecting supermajority in coin round (not decide). Hence, election will end with probability 1.
\item \textbf{FindOrder:} It decides round received of each event. Round received of event in round r is same as round number of famous witness of round r, if it can be seen by every famous witness of round r. Now, to determine consensus timestamp of event (say X in round r), then immediate ancestor of round r witness and descendant of X are found. Let us suppose that these events are E1, E2, E3, and so on upto En. Timestamp of all these events are taken which was assigned by their creator. Then median of these timestamp is assigned as a consensus timestamp of X.
\end{itemize}

Some important features of hashgraph are listed below:
\begin{itemize}
\item \textbf{Fairness:} Consensus timestamp is assigned to all events which ensures that their order is correct.
\item \textbf{Fast:} It has high throughput (number of transactions/second) as it works on gossip-about-gossip protocol. It ensures that propagation of events is fast in the network as it randomly selects a member and sends the message without any condition or boundation. It does not need to wait to get the votes of member to assign consensus timestamp.
\item \textbf{Virtual voting:} Every member maintains full history of transactions by maintaining DAG of events. No member is required to send it's vote to other member. This is because once hashgraph DAG is build, it is easy to know how a node will vote, since every node has all the information of what each node know and when they knew it. This information is used for finding order of transactions. Building a DAG on every node helps to achieve consensus on order of events independently. This saves a huge amount of bandwidth as node does not have to transfer their vote to other nodes. This is because other node has required amount of information of how a node will vote in election.
\item \textbf{Highly efficient:} No event is ever discarded in hashgraph. All the events are used when DAG is formed. Bandwidth requirement is also very less as only information related to transactions are transmitted and no votes are transmitted over the network.
\end{itemize}

\section{RELATED WORK}
Sankar et al. compared blockchain technologies on the basis of their consensus protocols [9]. Xu et al. compared blockchain on their advantages and disadvantages [10]. Lin et al. reviewed blockchain with the focus of security [11]. Application of different blockchain based DLT in the field of IOT was reported in [12]. Comparison of nano, a DAG based DLT, was done with blockchain on the surface by Beni et al. [13]. Schueffel et al. reviewed some DAG based DLT and blockchain in the literature.\\
Studies comparing blockchain and various DAG based DLT on extensive criteria and providing a big picture on current state of DLT could not be traced. This literature does that which helps to predict the direction in which DLT based technology is moving forth. 

\section{Criteria for Evaluation of Distributed Ledger Technologies}

By going over white papers of various DLTs discussed in previous sections[4-7], it was found that each of the DLTs tried to solve the shortcomings of present technologies by targeting set of features to distinguish itself from others. Although, targeting only set of features bring some disadvantages in these DLTs which are discussed in subsequent sections. Nowadays, new technologies brings unprecedented novelty and hence, it generates immense confusion. To bring innovation, one should have thorough knowledge of existing technologies, use case, strengths and weaknesses. Set of criteria mentioned in this literature will definitely help anyone to get a better view of current standings of various DLTs discussed in this paper. These criteria are listed below:
\begin{enumerate}
\item \textbf{Architecture:} how information is stored and arranged in the network.
\item \textbf{Transaction:} representation of transaction in the network.
\item \textbf{Consensus:} how transactions are validated in the network.
\item \textbf{Copyright:} whether rights given to use that technology
\item \textbf{Latency:} time taken by the network to validate a transaction.
\item \textbf{Byzantine Fault Tolerant (BFT):} Systems that tolerate failures that belong to Byzantine Generals’ Problem [2].
\item \textbf{Privacy:} whether DLT ensures privacy
\item \textbf{Fee:} Cost to submit a transaction  
\item \textbf{Throughput (tps):} rate of transactions administered by the network
\item \textbf{Security:} extent to which network grantees security 
\item \textbf{Fairness:} Transaction A will appear before transaction B in the consensus order, if A was created before B.
\item \textbf{Cost:} cost of participation
\item \textbf{Maturity:} present state of technology
\item \textbf{Setting:} whether technology is public or private
\item \textbf{Competition:} whether nodes are competing in the network to create block
\item \textbf{Scalable:} affect on throughput when size of network is increased
\item \textbf{Platform:} environment where technology is running
 
\end{enumerate}

\section{Comparative Evaluation of Distributed Ledger Technologies}
Quality criteria found in the previous section is used to evaluate DLTs on various grounds. The comparative results of various DLTs are shown in Table 1.\\
Blockchain has linked list data structure. Every element is known as a block and each block stores group (around 3500) of transactions. Sidechain extends this very idea and implements various linked list associated to each other. Tangle and hashgraph have DAG based data structure where each entity in the DAG represents a transaction in the former and each element in DAG is an event which store group of transactions in the latter.\\
If the main chain in blockchain branches off, then longest chain is taken as a main chain and others are pruned off. This is because it assumes that true nodes have more than 50\% of the computing power and hence, they will generate the longest chain even if malicious node try to manipulate the chain. DAG based chain continuously branches and converges later i.e., no block is ever stale and every block is used unlike blockchain.\\
Most of the DLTs are not fair unlike hashgraph. Fairness is a property where if a transaction A is generated earlier than transaction B, then A should appear before B in the consensus order of transactions. Only hashgraph is fair among various DLTs. Blockchain based DLTs are not fair as it is upto a miner who decides which transaction will go in which block.\\
Byzantine fault tolerant is a system which will eventually achieve consensus in the network even if some members are malicious and are trying to manipulate results or halt consensus. Blockchain is not Byzantine fault tolerant because nobody knows when complete consensus is achieved, rather probability of confidence increases as time passes. Whereas, Hashgraph is proven to be fully asynchronous Byzantine Fault Tolerance (aBFT) which is gold standard of security in the field of distributed systems [8].\\
Now, if we compare DLTs on the basis of participants, we find that blockchain has issuers who submits a transaction for validation and miners who maintains the integrity of the blockchain. There is coordinator node in sidechain which is present in each of the blockchain. IOTA has a coordinator which validates transaction and others are general users who submits the transaction for validation known as entities. Users of hashgraph are known to Swirlds and every member is known to every other member. Hence, all are allowed to validate transactions.\\
To maintain integrity of the network, consensus is required on state of the network. Blockchain uses POW and so does sidechain but with lower difficulty as it is permissioned network. Tangle uses coordinator node which issues transactions and, whichever transaction is referenced by it are considered confirmed. Hashgraph uses virtual voting to achieve consensus among members.

\begin{table}
\caption{Comparision of DLTs on various criteria}
\label{tab:example}
\centering
\tabcolsep=0.11cm
\begin{tabular}{l|p{1.5cm}|p{1.5cm}|p{1.5cm}|p{1.5cm}}
    \hline
    Criteria  &  Blockchain &  Sidechain &  Tangle &  Hashgraph\\
    \hline
    \hline
    Architecture  & Linked List Blocks  & Multiple linked list  & DAG  & DAG \\
    \hline
    Transaction  & grouped in block  & Grouped in block & separate entity & Grouped in Event \\
    \hline
    Consensus   & POW (SHA-256)  & POW (Ethash)  & POW (hashcard)  & Virtual voting \\
    \hline
    Copyright    & open source  & open source  &  open source & patented\\
    \hline
    Latency   & a few minutes  & a few minutes  & a few seconds  & less than 10 seconds\\
    \hline
    BFT    &No   & No  &  No & Yes\\
    \hline
    Privacy    & low  & high  & low  & low\\
    \hline
    Fee    &yes   & yes  &  no & no\\
    \hline
    Throughput  &5-20 tps   & limited by main chain  &500-800 tps  & 100,000 tps \\
    \hline
    Security    & high  & high  & high  & high\\
    \hline
     Fairness   &No   & No  &Not yet  &Yes \\
    \hline
     Cost   &high   & high  & low  &low \\
    \hline
    Maturity     &Many implementations  &Experimental   & Experimental   &Experimental  \\
    \hline
    Setting    & public  &  private and public  & private  & private\\
    \hline
    Competition   &  yes & yes  &  no & no\\
    \hline
    Scalable    & limited  &  yes & yes  & yes\\
    \hline
    Platform  & Ethereum, Bitcoin, etc  & Monax  & IOTA  & Hedra\\
    \hline
\end{tabular}
\end{table}

\section{RECENT DEVELOPMENTS}
 Facebook introduced libra cryptocurreny which will be launched in first half of 2020. Initially some big players will be allowed to join the network for validation such as visa and PayPal, although, Facebook has promised that they will migrate to permissionless network within 5 years.\\
It will be governed by Libra association and is built on secure scalable and reliable Libra blockchain which is open source. Rather than having blocks for storing transactions, it records the history of transactions in single data structure. Throughput of Libra will be around 1000 tps which is not as fast as hashgraph. But, throughput is reasonable as visa does around 1700 tps and for PayPal, it is around 193 tps. Consensus mechanism in Libra is known as LibraBFT. This protocol will work in environment where 1/3rd nodes can be Byzantine in the network. Libra is Byzantine fault tolerant, whereas hashgraph is asynchronous byzantine fault tolerant (aBFT, which is highest level of security) and the only DLT that is mathematically proven to be aBFT.

\section{Discussions}
Some of the issues with blockchain is its performance not meeting current needs, forks in the chain and transaction fee which might be greater than transaction itself. Sidechain tried to solve problem of scaling related to blockchain but it also has some limitations as it doesn't remove miners from the picture which require incentives. Hashgraph and tangle solved many problems of blockchain. For example, they both have better throughput than blockchain based DLT. Tangle removed the transaction fee from the network as user who submits the transaction for validation has to validate other transaction. Hashgraph uses gossip-about-gossip and virtual voting to increase throughput and efficiency of the network. It might be tempting to say tangle and hashgraph are better than blockchain based DLT, but there is much more to it. Two major blockchain platforms (Bitcoin and Ethereum) have been studied, but there are other blockchain platforms that have better performance. For example, Hyperledger (which runs in private setting) has throughput (around 700 tps) same as that of tangle. The reason why blockchain based DLT like bitcoin and ethereum are slow because they are public. In Public setting, anybody can join the network and start mining and can leave the network at their own will. Public blockchain completely removes the middleman (eg bank) from the picture. In public setting, no one can be trusted and hence, single party is now allowed to control the network. There is a choice between speed and decentralisation. Public blockchain chooses decentralisation over speed as it is the entire point of having public blockchain. Hence, throughput of the network is suppressed intentionally. This is done to reduce the number of branches or forks in the chain and allow block to completely propagate throughout the network, so that blockchain remains sync and fraudulent transactions are prevented. The reason why Hyperledger, tangle and hashgraph are fast because it is private and permissioned. Not anyone can join the private network. Only trusted parties are allowed to join and verify transactions. Therefore, it is kind of a `centralised system'. Private blockchain choses speed over complete decentralisation. This allows much greater throughput and efficiency while transactions are validated. Hashgraph takes private blockchain to next level by introducing consensus mechanism that increases throughput many times greater than any other DLT.\\
With the introduction of hashgraph followed by Libra, private DLT are picking up pace, it is likely that gap between public and private DLT will widen. Public blockchain is totally decentralised (and hence, no trust) and problem is to make consensus on transactions fast such that they are secure in the network where there is no mutual trust. Whereas, private blockchain is partially distributed (partial trust) and throughput is high but it is not open to public.\\
As private DLT is controlled by organisation, people might not have complete trust in the system as these corporations might act according to their interests. Bad actors can gain control if allowed to enter into the private network. Hashgraph has shown some progress as its network will achieve consensus if malicious nodes is less than 1/3rd of the total nodes in the network.\\
The main idea of blockchain was to remove trusted party completely from the picture is not completely met in private DLT. A perfect DLT should be best of both the worlds i.e., completely decentralised and high throughput.\\
Although, blockhain based DLT such as bitcoin is public rather than permissioned, but a recent study shows 74\% of the bitcoin miners are from China. That is, 74\% of the computing power resides in China which means that it is susceptible to 51\% attack (POW assumes that more than 50\% of the computing power is controlled by trusted nodes).\\

\section{CONCLUSION}
Huge potential is shown by Distributed Ledger Technologies that could be used in various industries. They are posing threat to the existence of present centralised third party organisations like banks which charge for their services.
In this paper, major DLTs were comprehensively reviewed and were compared on the basis of criteria mentioned in preceding section. Public blockchain such as bitcoin and ethereal were among the first DLTs that brought decentralisation, trust, security and low operational cost. However, they lacked the throughput required to meet needs of current services like PayPal among other shortcomings. DAG based solutions have claimed to solve many of challenges by public blockchain. IOTA is faster, but it still uses centralised coordinator for validation. Hashgraph on the other hand is much faster, fairer and secure than public blockchain. Both IOTA and hashgraph works in permissioned environment which is inconsistent with the idealogy of removing trusted third party.

\end{document}